\documentclass[twocolumn,showpacs,preprintnumbers,prb]{revtex4}
\usepackage{graphicx}
\usepackage{xcolor}
\usepackage{dcolumn}
\usepackage{amssymb}
\usepackage{bm}

\begin{document}

\title{Model anisotropic quantum Hall states}
\author{R.-Z. Qiu$^1$, F. D. M. Haldane$^2$, Xin Wan$^{3}$, Kun Yang$^{4}$, and Su Yi$^1$}
\affiliation{$^1$State Key Laboratory of Theoretical Physics, Institute of
Theoretical Physics, Chinese Academy of Sciences, Beijing 100190, China}
\affiliation{$^2$Department of Physics, Princeton University, Princeton NJ 08544-0708}
\affiliation{$^3$Zhejiang Institute of Modern Physics, Zhejiang University,
Hangzhou 310027, China}
\affiliation{$^4$National High Magnetic Field Laboratory and Department of Physics,
Florida State University, Tallahassee, Florida 32306, USA}

\begin{abstract}
Model quantum Hall states including Laughlin, Moore-Read and Read-Rezayi states are
generalized into appropriate anisotropic form. The generalized states are exact
zero-energy eigenstates of corresponding {\em anisotropic} two- or multi-body Hamiltonians,
and explicitly illustrate the existence of geometric degrees of in the fractional quantum Hall effect.
These generalized model quantum Hall states can provide a good description
of the quantum Hall system with anisotropic interactions.
Some numeric results of these anisotropic quantum Hall states are also presented.
\end{abstract}

\maketitle

\section{Introduction}
Variational wave functions play a fundamental role in our understanding of fractional quantum
Hall (FQH) effect, with the Laughlin wave function \cite{Laughlin} being the most
prominent example. New classes of quantum Hall trial wave functions have been proposed
as correlation functions in various conformal field theories,~\cite{MooreRead,ReadRezayi}
by generalizing the clustering properties of the wave functions,~\cite{simon07,wenwang} or
as Jack polynomials with a negative parameter and a matching root configuration.~\cite{JacksFQHS}
For a long time it has been understood that the Laughlin wave function, as well as other
FQH trial wave functions, contains {\em no} variational parameter.
This understanding is the consequence of our interest in searching for
{\em topologically distinct} quantum Hall wave functions;
geometry, as oppose to topology, is believed to be redundant.
Recently one of us~\cite{Metric} pointed out that such topological description of
FQH wave functions is not complete; in his geometrical description
the original Laughlin wave function is simply a member of a family of Laughlin
states, parameterized by a hidden (continuous) geometrical degree of freedom.
The family of the Laughlin states, with the geometrical factor as a variational parameter,
should provide a better description of the FQH effect in the presence of
either anisotropic effective mass or anisotropic interaction,
which are present in real materials.

The anisotropic FQH states are of present interest,
both theoretically~\cite{mulligan,QHEanis} and experimentally.~\cite{xia}
In Ref.~\onlinecite{QHEanis}, some of us studied the quantum Hall effects in a fast
rotating quasi-two-dimensional gas of polarized fermionic dipoles.
The fast rotation is equivalent to the high magnetic field according to the
Larmor theorem. And since $p$-wave interaction for the polarized fermions is typically
very small unless in the resonance regime, the only significant interaction is
the dipole-dipole interaction that could be tuned by adjusting the applied electric
and magnetic fields.
By tuning the direction of the dipole moment with respect to the $z$ axis,
the dipole-dipole interaction becomes anisotropic in the $x$-$y$ plane.
Specifically, the system contain
two parameters: the confining strength determined by the rotation frequency
and the tilt angle determined by the applied field.
Thus such systems have highly tunable anisotropic interactions,
and are ideal for studies of anisotropic FQH states.
Results in Ref.~\onlinecite{QHEanis} clearly indicate the {\em inadequacy}
of the known variational wave functions in the description of such states.

As a matter of fact, consideration of possible anisotropic FQH states has a long
history,~\cite{joynt,wexler,fogler} with variational
wave functions constructed that are straightforward modifications
of the original Laughlin wave function.~\cite{joynt,wexler}
Unlike the original Laughlin wave function and the states to be discussed here
however, these earlier anisotropic FQH states are {\em not}
exact eigenstates of special 2-body Hamiltonians.
Interest in such states was partially motivated by the observation of
{\em compressible} states with strong anisotropic transport properties
in high Landau levels (LLs).~\cite{lilly,du} In all these states, the rotation
symmetry is broken {\em spontaneously}. Very recently, an anisotropic
FQH state has been observed at $\nu=7/3$.~\cite{xia}
In this case the rotation symmetry is broken {\em explicitly} by an in-plane magnetic field,
whose direction dictates the transport anisotropy.
Generalizations of the FQH states to lattice models with zero net magnetic field
and full lattice translation symmetry is also called for to describe fractional quantum
anomalous Hall states and fractional topological insulators.~\cite{qi11}

Ref.~\onlinecite{Metric} pointed out the existence of a family of Laughlin states
that are zero energy ground states of a family of Hamiltonians consisting of
projection operators, which depend on a parameter called guiding center metric.
However the states themselves were not constructed explicitly, and only some of their
qualitative properties were mentioned briefly.
The main motivation of this work
is to explore the construction of a family of wave functions
{\em in closed form} with numerical comparisons to facilitate
the study of the geometrical aspects of anisotropic FQH states
as a result of anisotropic interaction in the planer geometry.

The rest of the paper is organized as follows. In Sec.~\ref{Background}
we introduce the anisotropic LL basis states using the unimodular transformation.
In Sec.~\ref{GFQH} we
focus on the recipe to generate
the anisotropic many-particle states due to the unimodular
transformation.
Section.~\ref{numerics} covers the numerical studies on
various properties of the family of Laughlin states,
including their density profile, pair correlation function,
and variational energy. We summarize the paper in Sec.~\ref{conclusion}.

\section{Anisotropic Landau level basis states}
\label{Background}

The original Laughlin wave function was most
easily written down on a disc, using the symmetric gauge in
which the single particle basis states are angular momentum eigenstates.
The key to explicitly constructing the
anisotropic Laughlin states proposed in Ref.~\onlinecite{Metric} (on a disc)
is the usage of a set of {\em anisotropic} LL basis states.
We will, however, continue to use the symmetric gauge,
in which the lowest LL (LLL) wave functions are holomorphic.
The same set of basis states appeared earlier in the consideration of
deformation of shape of quantum Hall liquids
in the context of Hall viscosity~\cite{Geometry},
although the wave functions were not explicitly given.
In the following
we generate them explicitly. In addition, we also use them
to construct the corresponding integer quantum Hall
wave function, as a warm-up for their later application to FQH states.

Let us start from the eigenvalues problem of
a two-dimensional charged particle subjected into
a uniform magnetic field ${\bm B}=B\hat{\bm z}$.
The one-body  Hamiltonian is given by
\begin{eqnarray}
H_0={\textstyle\frac{1}{2}} (m^{-1})^{ab}\pi_a\pi_b,
\quad \bm \pi = \bm p - e\bm A,
\end{eqnarray}
where we used the Einstein convention,
$m_{ab}$ is the cyclotron effective mass tensor,
$e>0$ is the charge of the particle,
${\bm A}$ is the vector potential of the uniform magnetic field ${\bm B}$,
and the dynamical momentum $\bm\pi$ satisfies
\begin{eqnarray}
[\pi_a,\pi_b]=i\epsilon_{ab}\hbar^2\ell^{-2}
\end{eqnarray}
with $\epsilon_{ab}=\epsilon^{ab}$ the 2D antisymmetric Levi-Civita symbol and
$\ell=\sqrt{\hbar/(eB)}$ the magnetic length.
In terms of a complex vector $\bm\omega$,
the effective mass tensor can be written as
\begin{eqnarray}
m_{ab}&=&m\ (\omega_a^*\omega_b + \omega_b^*\omega_a), \\
(m^{-1})^{ab}&=&m^{-1}(\omega^{a*}\omega^b + \omega^{b*}\omega^a).
\end{eqnarray}
we note that the assumption that it is isotropic in the $(x,y)$ Cartesian
coordinate system means
\begin{equation}
(\omega_x,\omega_y) = (\omega^x,\omega^y) =\frac{1}{\sqrt{2}}(1,i).
\end{equation}
The one-body Hamiltonian can now be expressed as
\begin{equation}
H_0 ={\textstyle \frac{1}{2}}\hbar \omega_c
\left( b^{\dagger}b + bb^{\dagger}\right).
\end{equation}
where $\hbar\omega_c$ = $ \hbar^2/m\ell^2$ is the cyclotron energy and
the Landau orbit ladder operators are given by
\begin{equation}
b = \hbar^{-1}\ell \left (\bm \omega \cdot \bm \pi\right ),
\quad b^\dagger = \hbar^{-1}\ell \left (\bm \omega^* \cdot \bm \pi\right ),
\end{equation}
with $[b,b^{\dagger}] = 1$. By introducing the ``guiding center'' coordinates ${\bm R}$,
\begin{eqnarray}
R^a=r^a+\hbar^{-1}\ell^2\epsilon^{ab}\pi_b,
\end{eqnarray}
which satisfy
\begin{eqnarray}
[R^a,R^b]=-i\epsilon^{ab}\ell^2,\quad [R^a,\pi_b]=0,
\end{eqnarray}
we could define a second set of ``guiding center'' ladder operators, $a$ and $a^{\dagger}$
that commute with $b$ and $b^{\dagger}$ (and hence with $H_0$)
\begin{eqnarray}
a=\ell^{-1}\left({\bm \omega}^*\cdot{\bm R}\right),
\quad a^\dagger=\ell^{-1}\left({\bm \omega}\cdot{\bm R}\right),
\quad [a,a^\dagger]=1.
\end{eqnarray}

To proceed further, let us
choose the vector potential of the uniform magnetic field
in the symmetric gauge,
\begin{eqnarray}
{\bm A}={\textstyle\frac{1}{2}}{\bm B}\times{\bm r}={\textstyle\frac{1}{2}}B(-y,x),
\end{eqnarray}
and define the complex coordinate,
\begin{equation}
z = \frac{\bm \omega \cdot \bm r}{\ell}.
\end{equation}
Let us note that for the case of isotropic mass the complex coordinate is explicitly given by
\begin{equation}
z = \frac{x+iy}{\sqrt{2}\ell},
\end{equation}
which is not the most standard definition in the literature.
Then we could express the Landau orbit ladder operators as
(neglecting the trivial $-i$ and $i$),
\begin{equation}
b = {\textstyle \frac{1}{2}}z + \partial_{z*} ,\quad  b^{\dagger} =
{\textstyle \frac{1}{2}}z^* - \partial_{z},
\end{equation}
where $\partial_zf(z,z^*) $ is the partial derivative $
\partial f/\partial z\vert_{z^*}$, \text{etc}, and
the ``guiding center'' ladder operators as
\begin{equation}
a = {\textstyle \frac{1}{2}}z^* + \partial_{z} ,\quad  a^{\dagger} =
{\textstyle \frac{1}{2}}z - \partial_{z^*} .
\end{equation}

The Hamiltonian $H_0$ has a rotational symmetry generated by
\begin{equation}
L_0 = a^{\dagger}a - b^{\dagger}b, \quad [L_0,H_0] =0,
\end{equation}
where
\begin{eqnarray}
&&[L_0,x] = i y,\quad [L_0,y] =-i x, \\
&&[L_0,p_x] = i p_y,\quad [L_0,p_y] =-i p_x.
\end{eqnarray}
Simultaneous diagonalization of $H_0$ and $L_0$
allows a complete orthonormal basis of one particle states $|\psi_{nm}\rangle$
which used to be constructed as
\begin{eqnarray}
&&H_0|\psi_{nm}\rangle = (n+{\textstyle\frac{1}{2}})\hbar
\omega_c|\psi_{nm}\rangle, \\
&&L_0|\psi_{nm}\rangle = (m-n)|\psi_{nm}\rangle, \\
&&|\psi_{nm}\rangle =
\frac {(a^{\dagger})^m(b^{\dagger})^n}{\sqrt{m!n!}}|\psi_{00}\rangle,\label{common}\\
&& a|\psi_{00}\rangle = b|\psi_{00}\rangle = 0.
\end{eqnarray}
This is, of course, the commonly used basis states associated with the symmetric gauge.

However, it is possible to construct a set of closely related but
{\em different} set of basis states.
First let us write
\begin{eqnarray}
L_0 &=&  L + \bar L,\\
L  &=&  {\textstyle\frac{1}{2}}(b^{\dagger}b + bb^{\dagger}), \quad H_0  =  \hbar
\omega_cL ,\\
\bar L &=& {\textstyle\frac{1}{2}}(a^{\dagger}a + aa^{\dagger}).
\end{eqnarray}
The conventional basis is then just the set of the simultaneous eigenstates of
$H_0$ and $\bar L$,
\begin{equation}
\bar L |\psi_{nm} \rangle =
(m+{\textstyle\frac{1}{2}})|\psi_{nm}\rangle.
\end{equation}
We may write  $\bar L $ = $\bar L(g^0)$, where
\begin{equation}
\bar L(g) =  \frac{1}{2\ell^2}g_{ab}R^aR^b,
\end{equation}
and $g_{ab}$ is a positive-definite Euclidean metric tensor with $\det g$ = 1.
Note that $\bar L(g^0)$ uses the
``Galilean metric''  derived from the effective
mass tensor,
\begin{eqnarray}
m_{ab}=m(\omega^*_a\omega_b+\omega^*_b\omega_a)=mg^0_{ab}.
\end{eqnarray}
However, as has been stressed by one of us\cite{halvisc,Metric}, that
before specifying the two-body interactions,
there is no fundamental reason to choose a guiding-center basis that
is adapted to the shape of the Landau orbits.
There is a more general family of bases,
parameterized by a complex number $\gamma$ with $|\gamma| < 1$,
\begin{eqnarray}
\bar \omega_a = (1-|\gamma|^2)^{-1/2}(\omega_a + \gamma^* \omega_a^*),
\end{eqnarray}
that determines a unimodular metric $g_{ab}(\gamma)$,
\begin{eqnarray}
\frac{1}{1-\gamma^*\gamma}\left(\begin{array}{cc}
(1+\gamma)(1+\gamma^*) & i(\gamma-\gamma^*)\\
i(\gamma-\gamma^*) & (1-\gamma)(1-\gamma^*) \\
\end{array}\right)\nonumber
\end{eqnarray}
according to
\begin{eqnarray}
\bar \omega_a^*\bar \omega_b  = {\textstyle\frac{1}{2}}\left ( g_{ab}(\gamma) +
i\epsilon_{ab}\right ),
\end{eqnarray}
and then $ g_{ab}(\gamma)=\bar \omega_a^*\bar \omega_b+\bar \omega_a\bar \omega_b^*$.
We can then construct a basis of  eigenstates $|\psi_{nm}(\gamma)\rangle$  of $H_0$ and
$\bar L(\gamma)$  $\equiv$ $\bar L(g(\gamma))$,
which is accomplished by defining a new
set of ``guiding center'' ladder operators, also considered in Ref.~\onlinecite{Geometry},
\begin{equation}
a_\gamma =\frac{{\bar {\bm\omega}^*}\cdot{\bm R}}{\sqrt{2}\ell} ,\quad
 a_\gamma^\dagger =
\frac{{\bar {\bm\omega}}\cdot{\bm R}}{\sqrt{2}\ell},
\end{equation}
and explicitly as
\begin{eqnarray}
\left(\begin{array}{c}
a_\gamma \\
a_\gamma^\dag \\
\end{array}\right)
=\frac{1}{\sqrt{1-\gamma^*\gamma}}
\left(\begin{array}{cc}
1 & \gamma  \\
\gamma^*  & 1 \\
\end{array}\right)
\left(\begin{array}{c}
a \\
a^\dag \\
\end{array}\right).
\end{eqnarray}
This is a Bogoliubov (or ``squeezing'')
transformation that preserves the commutation relation
$[a_{\gamma},a^{\dagger}_{\gamma}]$ = 1.
Then the new simultaneous eigenstates of $H_0$ and ${\bar L}_{0}(\gamma)$ are
given by
\begin{eqnarray}
&&|\psi_{nm}(\gamma)\rangle
=\frac{(b^\dag)^n(a^{\dagger}_{\gamma})^m}{\sqrt{n!m!}}
|\psi_{00}(\gamma)\rangle, \\
&& a_{\gamma}|\psi_{00}(\gamma)\rangle =
b|\psi_{00}(\gamma)\rangle  = 0.
\end{eqnarray}
Note that $|\psi_{nm}(0)\rangle$ is the conventional basis
$|\psi_{nm}\rangle$ in Eq.~(\ref{common}).

Since the model quantum Hall states are often defined in the lowest
Landau level (LLL)  ($n=0$), it is useful to examine the wave functions
$\psi_{0m}(\gamma,\bm r)$ = $\langle \bm r |
\psi_{0m}(\gamma)\rangle$.
First of all, from the LLL property $b\psi_{0m}(\gamma, \bm r)=0$,
we have
\begin{eqnarray}
\psi_{0m}(\gamma, \bm r)=f(z)\psi_{00}(0,\bm r)
\end{eqnarray}
with
\begin{eqnarray}
\psi_{00}(0,\bm r)=(2\pi)^{-1/2}e^{-|z|^2/2}.
\end{eqnarray}
Second from
$a_{\gamma}\psi_{00}(\gamma,\bm r)=0$,
we immediately obtain
\begin{eqnarray}
\psi_{00}(\gamma,\bm r)=\lambda^{1/2}e^{-\frac{1}{2}\gamma z^2}
\psi_{00}(0,\bm r)
\end{eqnarray}
with
\begin{eqnarray}
\lambda=\sqrt{1-|\gamma|^2}.\label{lambda}
\end{eqnarray}
Next let us note that $\psi_{0m}(\gamma, \bm r)$,
\begin{eqnarray}
\frac{\left[
\textstyle{\frac{1}{2}}z-\partial_{z^*}+
\gamma^*\left(
\textstyle{\frac{1}{2}}
z^*+\partial_z\right)
\right]^m}{\lambda^m\sqrt{m!}}
\psi_{00}(\gamma,\bm r),\nonumber
\end{eqnarray}
could be simplified by virtue of the property that
$b\psi_{00}(0, \bm r)=0$, or
\begin{eqnarray}
\partial_{z^*}\left[\psi_{00}(0,\bm r)f(z)\right]
=(-\textstyle{\frac{1}{2}}z)\psi_{00}(0,\bm r)f(z)
\end{eqnarray}
and $a\psi_{00}(0, \bm r)=0$, or
\begin{eqnarray}
\partial_{z}\left[\psi_{00}(0,\bm r)f(z)\right]
=\psi_{00}(0,\bm r)(\partial_z-\textstyle{\frac{1}{2}}z^*)f(z)
\end{eqnarray}
and
\begin{eqnarray}
\partial_{z}\left[e^{-\frac{1}{2}\gamma z^2} f(z)\right]
=e^{-\frac{1}{2}\gamma z^2}(\partial_{z}-\gamma z)f(z).
\end{eqnarray}
Thus we obtain
\begin{eqnarray}
\psi_{0m}(\gamma,\bm r)
&=&\psi_{00}(\gamma,\bm r)
\frac{\lambda^{m}}{\sqrt{m!}}(z+z_0^2\partial_z)^m
1\label{form}\\
&=&
\psi_{00}(\gamma,\bm r)
\frac{\lambda^{m}}{\sqrt{m!}}
W_m(z,z_0^2)
\end{eqnarray}
with
\begin{eqnarray}
z_0^2=\gamma^*/\lambda^2\label{z0}
\end{eqnarray}
and $W_m(z,z_0^2)$ are the polynomials defined by $W_0(z,z_0^2)$ = 1,
and the recursion relations,
\begin{eqnarray}
W_{m+1}(z,z_0^2)  &=& (z + z_0^2\partial_z)W_m(z,z_0^2),\\
W_{m}(z,z_0^2)  &=& \partial_zW_{m+1}(z,z_0^2)/(m+1).
\end{eqnarray}
The first few of $W_m(z,z_0^2)$ are given by
\begin{eqnarray}
W_{0}(z,z_0^2)&=&1,\nonumber\\
W_{1}(z,z_0^2)&=&z,\nonumber\\
W_{2}(z,z_0^2)&=&z^2+z_0^2,\nonumber\\
W_{3}(z,z_0^2)&=&z^3+3z_0^2z,\nonumber
\end{eqnarray}
and the general solutions are given by
\begin{eqnarray}
W_{2k}(z,z_0^2) &=& k!(2z_0^2)^kL^{(-\frac{1}{2})}_k
(-z^2/2z_0^2)
, \\
W_{2k+1}(z,z_0^2)& =& k!
(2z_0^2)^kzL^{(\frac{1}{2})}_k(-z^2/2z_0^2) ,
\end{eqnarray}
where $L_k^{(\alpha)}(x)$ are generalized Laguerre polynomials defined by
\begin{eqnarray}
L^{(\alpha)}_k(x)=\sum_{i=0}^k
\frac{\Gamma(\alpha+k+1)}{\Gamma(\alpha+i+1)\Gamma(k-i+1)}
\frac{(-x)^i}{i!}.
\end{eqnarray}
These expression do satisfy the reclusion relation, which could be proved
by using the following useful identities for $L_k^{(\alpha)}(x)$,
\begin{eqnarray}
L^{(\alpha)}_k(x)&=&L^{(\alpha+1)}_k(x)-L^{(\alpha+1)}_{k-1}(x),\\
\frac{d}{dx}
L^{(\alpha)}_k(x)&=&-L^{(\alpha+1)}_{k-1}(x),\\
xL^{(\alpha+1)}_{k-1}(x)&=&(k+\alpha)L^{(\alpha)}_{k-1}(x)-kL^{(\alpha)}_{k}(x).
\end{eqnarray}
Note that for $|z|^2 \gg |z^2_0|$, $W_m(z,z^2_0)\rightarrow z^m$.

\begin{figure}[tbp]
\centering
\includegraphics[width=3.2in]{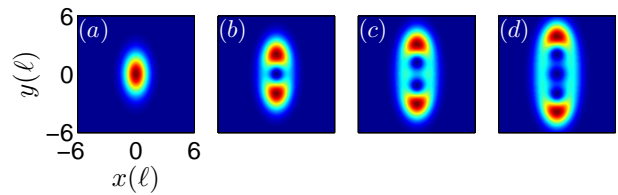}
\caption{(Color online) The density profiles of the
generalized LLL wave functions $\psi_{0m}(\gamma=1/2,\bm r)$
with $m=0$ (a), $m=1$ (b), $m=2$ (c) and $m=3$ (d).}
\label{AnisLLL1}
\end{figure}

\begin{figure}[tbp]
\centering
\includegraphics[width=3.2in]{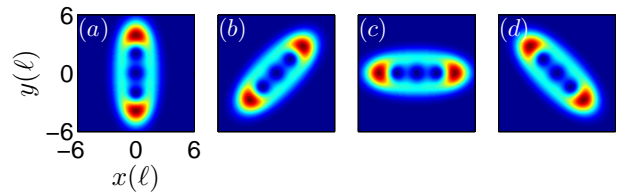}
\caption{(Color online) The density profiles of $\psi_{03}(\gamma,\bm r)$
 with $\gamma=e^{i\phi}/2$ for
$\phi=0$ (a), $\phi=\pi/2$ (b), $\phi=\pi$ (c) and $\phi=3\pi/2$ (d).
Note that the zeros are localized at $z=0$ and $\pm i\sqrt{3}z_0$.}
\label{AnisLLL2}
\end{figure}

The densities $|\psi_{0m}(\gamma,\bm r)|^2$
defined by the single-particle orbitals have
some noteworthy features.
First, the density profile is anisotropic for $\gamma\neq 0$.
Take a simple example like $\psi_{00}(|\gamma|,{\bm r})$,
\begin{eqnarray}
|\psi_{00}(|\gamma|,{\bm r})|^2\propto\exp\left[-\frac
{(1+|\gamma|)x^2+(1-|\gamma|)y^2}{2\ell^2}\right].\nonumber
\end{eqnarray}
For this Gaussian wave packet, the width along $y$-axis is larger than
the width along $x$-axis, so non-zero $|\gamma|$ causes stretching the wave function
along some direction.
Second, $|\psi_{0m}(\gamma,\bm r)|^2$ has $m$ zeros as can be seen from
Fig.~\ref{AnisLLL1} in which we plot the
density profiles $|\psi_{0m}(1/2,{\bm r})|^2$ for $m=0,1,2,3$.
These zeros are roots of  $W_m(z,z_0^2)$,
and they are aligned along
the stretched direction.
In the limit of $\gamma\rightarrow0$, they collapse to a multiple root
at the origin.
Third, the stretching direction is determined by the phase of the complex number
$\gamma$, ${\rm arg}(\gamma)$, which is easy to understand by noting
that $W_3(z,z_0^2)=0$ gives rise to three roots, $0$ and $\pm i\sqrt{3}z_0$.
In Fig.~\ref{AnisLLL2},
we plot $|\psi_{03}(e^{i\phi}/2,{\bm r})|^2$ with $\phi=0,\pi/2,\pi$ and $3\pi/2$.
Thus we realize that
the density profile $|\psi_{0m}(\gamma,{\bm r})|^2$ is equivalent to
rotating clockwise the density profile $|\psi_{0m}(|\gamma|,{\bm r})|^2$ by
${\rm arg}(\gamma)/2$.

To proceed further, let us define here the non-interacting basis
states to prepare for the many-body problem.
For bosonic system, the non-interacting basis state in the generalized LLL
is given by
\begin{eqnarray}
{\mathcal M}_{\mu}(\gamma)
\propto
\mathop{\rm perm}
\left[\psi_{0\mu}(\gamma,\bm r)\right],\label{nonintb}
\end{eqnarray}
where $\mu=[\mu_1,\mu_2,\ldots,\mu_N]$ is a non-decreasing partition and
$\mathop{\rm perm}[\psi_{0\mu}(\gamma,\bm r)]$ is the permanent of the square
matrix whose matrix elements are
$\psi_{\mu_i}(\gamma,\bm r_j)$ $(i,j=1,\ldots,N)$.
For the fermionic system, the non-interacting basis state
is defined by a Slater determinant,
\begin{eqnarray}
{\mathfrak{sl}}_{\mu}(\gamma)=\det\left[\psi_{0\mu}(\gamma,\bm r)\right],\label{nonintf}
\end{eqnarray}
where $\mu$ is a decreasing partition and $\det$ denotes
the matrix determinant.

As a special example, the Slater
determinant ${\mathfrak{sl}}_{\mu_0}(\gamma)$ with the partition
$\mu_0=[N-1,N-2,\ldots,1,0]$ is the generalized
$N$-particle IQH state, whose (unnormalized) expression
could be simplified as
\begin{eqnarray}
\Psi_{I}(\gamma)=\prod_{i<j}(z_i-z_j)\prod_i\psi_{00}(\gamma,\bm r_i),
\label{gIQH}
\end{eqnarray}
as lower-degree polynomials must vanish due to the  Pauli exclusion principle
(this is generally not the case for other partitions, which is the key to
understanding the anisotropic FQH states).

The generalized IQH state can also be viewed as a coherent
superposition of the isotropic IQH state (or simple
Vandermonde determinant state) and its edge states
with angular momentum differences being multiples of
$2\hbar$. This is achieved by expanding the exponential factor
in Eq.~(\ref{gIQH}). From the viewpoint of the superposition,
this is consistent with the numerical result in Ref.~\onlinecite{QHEanis}.

\section{Anisotropic many-particle states}
\label{GFQH}

In this section, we first consider the two-particle problem and
then simplify the interaction Hamiltonian projected into the LLL in Subsec.~\ref{two}.
More importantly, we present the recipe to find
the anisotropic counterpart of a LLL state and elaborate it in Subsec.~\ref{many}.
We specifically consider the anisotropic counterpart
of the prominent FQH states such as the Laughlin state in Subsec.~\ref{AFQH}.

\subsection{Two-particle problem and the projected two-body interaction Hamiltonian}
\label{two}

If the interaction is dominated by the energy gap $\hbar\omega_c$ between
the lowest and second Landau level, the interaction Hamiltonian
could be projected into the lowest Landau level.
In terms of non-commuting ``guiding center'' coordinates ${\bm R}_i$,
the projected two-body interaction Hamiltonian we consider here is given by \cite{Metric}
\begin{eqnarray}
H_{\rm int}(\{V_m\},\gamma)
=\sum_{k} V_m h_m(\gamma),
\end{eqnarray}
where $V_m$ is the anisotropic Haldane's pseudopotential and the projection
operator $h_m(\gamma)$ is given by
\begin{eqnarray}
h_m(\gamma) &=& \int \frac{d^2\bm q\ell^2}{2\pi} v_m(\bm q,g(\gamma))
\sum_{i < j} e^{i\bm q \cdot (\bm  R_i-\bm R_j)}, \\
v_m(\bm q,g) &=&
L_m(g^{ab}q_aq_b\ell^2)\exp (-{\textstyle\frac{1}{2}}g^{ab}q_aq_b\ell^2),
\end{eqnarray}
where $L_m(x)=L^{(\alpha=0)}_m(x)$ are the Laguerre polynomials.
In addition,
the total angular momentum associated with the guiding center
is given by
\begin{eqnarray}
{\bar L}(\gamma)&=&\frac{1}{2\ell^2}\sum_ig_{ab}R^{ia}R^{ib}.
\end{eqnarray}

For two-particle, the LLL states that simultaneously diagonalize
${\bar L}(\gamma)$
and the interaction $H_{\rm int}(\{V_k\},\gamma)$ are completely fixed by symmetry.
To clarify this, let us note that
\begin{eqnarray}
{\bar L}(\gamma)
&=&a^\dag_{1\gamma}a_{1\gamma}+a^\dag_{2\gamma}a_{2\gamma}+1
\nonumber\\
&=&a^\dag_{12\gamma}a_{12\gamma}+A^\dag_{12\gamma}A_{12\gamma}+1
\end{eqnarray}
with
$a_{12\gamma}=(a_{1\gamma}-a_{2\gamma})/\sqrt{2}$ and
$A_{12\gamma}=(a_{1\gamma}+a_{2\gamma})/\sqrt{2}$, and
the only term containing operator
in the interaction Hamiltonian, $\bm  R_1-\bm R_2$ , are given by
\begin{eqnarray}
R^{x}_1-R^{x}_2&=&\frac{\ell}{\lambda}
\left[(1-\gamma^*)a_{12\gamma}+(1-\gamma)a_{12\gamma}^\dag\right]
\nonumber\\
R^{y}_1-R^{y}_2&=&\frac{\ell}{i\lambda}
\left[(1+\gamma)a^\dag_{12\gamma}-(1+\gamma^*)a_{12\gamma}\right]
\nonumber
\end{eqnarray}
and contains only $a_{12\gamma}$ and $a^\dag_{12\gamma}$.
Therefore, we could expand ${\bar L}(\gamma)$ and $H_{\rm int}(\{V_k\},\gamma)$
in the orthonormal basis
\begin{eqnarray}
&&|Mm\rangle_\gamma=
\frac{(A^\dag_{12\gamma})^M(a^\dag_{12\gamma})^m}{\sqrt{M!m!}}
|00\rangle_\gamma, \\
&&a_{1\gamma}|00\rangle_\gamma=a_{2\gamma}|00\rangle_\gamma=0,
\end{eqnarray}
and find that
\begin{eqnarray}
&&H_{\rm int}(\{V_k\},\gamma)|Mm\rangle_\gamma=V_m|Mm\rangle_\gamma,
\label{inteigen}\\
&&{\bar L}(\gamma)|Mm\rangle_\gamma=(M+m+1)|Mm\rangle_\gamma.
\end{eqnarray}

Thus the projected operator $h_m(\gamma)$ for many-particle system
could be rewritten as
\begin{eqnarray}
h_m(\gamma)=\sum_{i<j}\left(\sum_M|Mm\rangle_\gamma{}_\gamma M m|\right)(i,j)
\end{eqnarray}
and in second quantized form as
\begin{eqnarray}
\frac{1}{2}
\sum_M
\sum_{m_1m_2m_3m_4}&&{}_\gamma\langle m_1,m_2|mM\rangle_\gamma
{}_\gamma\langle m M|m_3,m_4\rangle_\gamma
\nonumber\\
&&g^\dag_{m_1}g^\dag_{m_2}
g_{m_4}g_{m_3},
\end{eqnarray}
with $|m_1,m_2\rangle=|\psi_{0,m_1}(\gamma)\rangle\otimes
|\psi_{0,m_2}(\gamma)\rangle$ and
$g^\dag_{m}$ being the creation operator that creates a particle in state
$|\psi_{0m}(\gamma)\rangle$.

For the rotationally-invariant case, $\gamma=0$,
\begin{eqnarray}
\Psi_{Mm}(0,{\bm r}_1,{\bm r}_2)
=\frac{(z_1+z_2)^M}{\sqrt{2^MM!}}
\frac{(z_1-z_2)^m}{\sqrt{2^mm!}}
\prod_{i=1}^2\psi_{00}(0,{\bm r}_i)
\nonumber
\end{eqnarray}
and the translation into the Heisenberg state is
\begin{eqnarray}
|Mm\rangle_\gamma
=\frac{(a^\dag_1+a^\dag_2)^M}{\sqrt{2^MM!}}
\frac{(a^\dag_1-a^\dag_2)^m}{\sqrt{2^mm!}}
|00\rangle_\gamma.
\end{eqnarray}
Therefore we find the anisotropic deformation is now easily achieved by
the simple replacement
\begin{eqnarray}
a_i^\dag\rightarrow a^\dag_{i\gamma},
\end{eqnarray}
and the vacuum state $|\Psi_0(0)\rangle\rightarrow|\Psi_0(\gamma)\rangle $ that
satisfy $a_{i\gamma}|\Psi_0(\gamma)\rangle=0$.
In the next subsection, we shall utilize this idea to consider the anisotropic deformation
of a general LLL wave function.

\subsection{Anisotropic many-particle states}
\label{many}

The general form of a rotationally-invariant $N$-particle LLL wave function in the
symmetric gauge is
\begin{equation}
\Psi(\bm r_1,\ldots ,\bm r_N) = F(z_1,\ldots, z_N)\prod_i
\psi_{00}(\bm r_i),
\end{equation}
where $F(z_1,\ldots ,z_N)$ is a homogeneous multivariate polynomial
with $F(\lambda z_1,\ldots \lambda z_N)=\lambda^MF(z_1,\ldots,
z_N)$ and $\bar L(0)|\Psi\rangle$ = $(M + {\textstyle\frac{1}{2}}N)|\Psi\rangle$.
The translation to the Heisenberg picture yields
\begin{eqnarray}
&&|\Psi(\gamma=0)\rangle = F(a_1^{\dagger},\ldots ,a^{\dagger}_N)|\Psi_0(\gamma=0)\rangle,
\nonumber\\
&&a_i|\Psi_0(\gamma=0)\rangle = b_i|\Psi_0(\gamma=0)\rangle = 0.
\end{eqnarray}
The deformation of
this Heisenberg state becomes
\begin{eqnarray}
&&|\Psi(\gamma)\rangle = F(a_{\gamma, 1}^{\dagger},\ldots ,a_{\gamma, N}^\dagger)
|\Psi_0(\gamma)\rangle,\nonumber
\\
&&
a_{\gamma,i}|\Psi_0(\gamma)\rangle = b_i|\Psi_0(\gamma)\rangle = 0.
\end{eqnarray}
The Schrodinger wave function for this state $|\Psi(\gamma)\rangle$ will have the form,
\begin{equation}
\Psi(\gamma,\bm r_1,\ldots,\bm r_N) =
F_{\gamma}(z_1,\ldots z_N) \prod_i\psi_{00}(\bm r_i).
\end{equation}

We now need the formal construction of $F_{\gamma}(\{z_i\})$, which is still
holomorphic, but no longer a polynomial.
Using the homogeneity of the polynomial $F$,
the wave function is given by
\begin{eqnarray*}
\lambda^{-M}F(\{{\textstyle\frac{1}{2}}z_i - \partial_{z^*_i}
+ \gamma^*({\textstyle\frac{1}{2}}z^*_i + \partial_{z_i})\})
\prod_i\lambda^{\frac{1}{2}}e^{-\frac{1}{2}\gamma z_i^2} \psi_{00}(\bm r_i).
\end{eqnarray*}
Using the property $a_i\psi_{00}(\bm r_i)$ = 0, or
\begin{equation}
\partial_z\psi_{00}(\bm r) f(z,z^*) =
\psi_{00}(\bm r)(\partial_z - {\textstyle\frac{1}{2}}z^*)f(z,z^*),
\end{equation}
and $b_i\psi_{00}(\bm r_i)$ = 0,
or
\begin{equation}
\partial_{z*}\psi_{00}(\bm r) f(z,z^*) =
\psi_{00}(\bm r)(\partial_{z*} - {\textstyle\frac{1}{2}}z)f(z,z^*),
\end{equation}
the holomorphic part $F_{\gamma}(\{z_i\})$  of the wave function becomes
\begin{eqnarray*}
\lambda^{-M}F(\{z_i  + \gamma^*\partial_{z_i}\}) \prod_i\lambda^{\frac{1}{2}}
e^{-\frac{1}{2}\gamma z_i^2}.
\end{eqnarray*}
Using the following equality,
\begin{equation}
\partial_z \left(e^{-\frac{1}{2}\gamma z_i^2}f(z)\right)=
e^{-\frac{1}{2}\gamma z_i^2}\left (\partial_z - \gamma z\right )
f(z),
\end{equation}
$F_{\gamma}$ turns into,
\begin{eqnarray*}
\left ( \prod_i \lambda^{\frac{1}{2}}e^{-\frac{1}{2}\gamma z_i^2}\right )
\lambda^M F(\{z_i + z_0^2\partial z_i\}) 1,
\end{eqnarray*}
which is easy to understand by noting Eq.~(\ref{form}).
Using the homogeneity of $F$, we obtain  the final result
\begin{equation}
F_{\gamma}(\{\lambda^{-1} z_i\}) =
\left ({\textstyle \prod_i}\lambda^{\frac{1}{2}}e^{-\frac{1}{2} \gamma  z_i^2}\right )
F(\{z_i + z_0^2\partial_{z_i}\}) 1.
\end{equation}
where $\lambda$ and $z^2_0$  are fixed by $\gamma$
through Eqs.~(\ref{lambda}) and (\ref{z0}).
This expression no longer requires homogeneity of $F(\{z_i\})$,
and is quite general for the deformation of the holomorphic part of
a LLL wave function.

Specially, the anisotropic counterpart of the rotationally-invariant
non-interacting basis states (\ref{nonintb},\ref{nonintf})
\begin{eqnarray}
{\mathcal M}_\mu(\gamma=0),\quad {\mathfrak{sl}}_\mu(\gamma=0),
\end{eqnarray}
are
\begin{eqnarray}
 {\mathcal M}_\mu(\gamma),\quad {\mathfrak{sl}}_\mu(\gamma).
\end{eqnarray}
Therefore,
if we could expand the many-body state $\Psi(\gamma=0,{\bm r}_1,\ldots,{\bm r}_N)$
in the non-interacting basis states ${\mathcal M}_\mu(0)$ (\ref{nonintb}) or
${\mathfrak{sl}}_\mu(0)$ (\ref{nonintf}),
\begin{eqnarray}
 \sum_{\mu}c^{(b)}_\mu{\mathcal M}_\mu(0),\quad\sum_{\mu}c_\mu{\mathfrak{sl}}_\mu(0),
\end{eqnarray}
its anisotropic counterpart $\Psi(\gamma,{\bm r}_1,\ldots,{\bm r}_N)$ would be
\begin{eqnarray}
\sum_{\mu}c^{(b)}_\mu{\mathcal M}_\mu(\gamma),\quad
\sum_{\mu}c_\mu{\mathfrak{sl}}_\mu(\gamma).
\end{eqnarray}
Here we find the coefficients $c_\mu$ and $c^{(b)}_{\mu}$ are independent
of $\gamma$.

The $\gamma$-independence of the coefficients results in
$\gamma$-independence of
the occupation number for each orbital
$$n_m=\langle \Psi^q_L(\gamma)|g_m^\dag g_m|\Psi^q_L(\gamma)\rangle.$$
This also suggests that the entanglement spectrum,~\cite{ES}
which encodes the topological properties of the state,
is also invariant, if the appropriate cut in the
space of the total angular momentum associated with guiding center is chosen.
We see that topological properties of a FQH state are built into
the coefficients (relative weight), which are
manifested, e.g., in the entanglement spectrum calculation.
On the other hand, geometrical properties of the anisotropic FQH states
are encoded into the deformation of the non-interacting basis.
Nevertheless, geometrical information can be revealed by exploring
the properties of the family of variational wave functions and
the corresponding variational energies; this point will be illustrated
by examples in Sec.~\ref{numerics}.

\subsection{Anisotropic FQH states}
\label{AFQH}

In this subsection, let us consider some prominent variational functions such
as the Laughlin wave function, Moore-Read state and so on.

The $\nu=1/q$ Laughlin wave function is given by
\begin{eqnarray}
\Psi^q_L(\gamma=0)=
\prod_{i<j}(z_i-z_j)^q\prod_{i}\psi_{00}(0,{\bm r}_i),
\end{eqnarray}
and its anisotropic counterpart is given by
\begin{eqnarray}
\Psi^q_L(\gamma)=\prod_{i}\psi_{00}(\gamma,{\bm r}_i)
\prod_{i<j}\left[
z_i-z_j+z_0^2(\partial_{z_i}-\partial_{z_j})\right]^q 1.\nonumber\\
\end{eqnarray}
The anisotropic Laughlin states $\Psi^q_L(\gamma)$ are the
(minimum total $\bar L(\gamma)$) zero-energy ground state of
the Haldane's pseudopotential Hamiltonian $H_{\rm int}(\{V_m\},\gamma)$ with
\begin{eqnarray}
V_{m<q}>0 \mbox{ and } V_{m\geq q}=0.
\end{eqnarray}
To prove this, we may consider the diagonalization of the
projected interaction Hamiltonian matrix
and note the $\gamma$-independence of the matrix elements
\begin{eqnarray*}
&&\langle{\mathfrak{sl}}_{\mu}(\gamma)|H_{\rm int}(\{V_m\},\gamma)
|{\mathfrak{sl}}_{\mu'}(\gamma)\rangle, \;{\rm or} \\
&&\langle{\cal M}_{\mu}(\gamma)|H_{\rm int}(\{V_m\},\gamma)
|{\cal M}_{\mu'}(\gamma)\rangle,
\end{eqnarray*}
which results from the $\gamma$-independence of ${}_\gamma\langle m_1,m_2|mM\rangle_\gamma$.

For two particles, the anisotropic counterpart of the $\nu=1/q$ Laughlin state
\begin{eqnarray}
\Psi_{0q}(0,\bm r_1,\bm r_2)=
{(z_1-z_2)^q}\Psi_{00}(0,{\bm r}_1,{\bm r}_2)\label{2Ls}
\end{eqnarray}
is given by
\begin{eqnarray}
&&\Psi_{0q}(\gamma,\bm r_1,\bm r_2)\nonumber\\
&=&\psi_{00}(\gamma,{\bm r}_1)\psi_{00}(\gamma,{\bm r}_2)
\left[z_1-z_2+z_0^2(\partial_{z_1}-\partial_{z_2})\right]^q1,\nonumber\\
&=&\psi_{00}(\gamma,{\bm r}_1)\psi_{00}(\gamma,{\bm r}_2)
(z_{12} + 2z_0^2\partial_{z_{12}} )^q1,\nonumber\\
&=& \psi_{00}(\gamma,{\bm r}_1)\psi_{00}(\gamma,{\bm r}_2)
W_q(z_{12},2z_0^2),\label{a2Ls}
\end{eqnarray}
with $z_{12}=z_1-z_2$ and $\partial_{z_{12}}={\textstyle\frac{1}{2}}
\left (\partial_{z_1} - \partial_{z_2}\right )$.
Among these states, the straightforward (anisotropic) generalization
of the $\nu=1/2$ bosonic Laughlin state is $\Psi_{02}(\gamma,\bm r_1,\bm r_2)$,
in which the double zero of the undeformed state
at $z_1=z_2$ is split into zeroes at
$z_1-z_2$ = $\pm i\sqrt{2}z_0$. While for the $\nu=1/3$ fermionic Laughlin state
$\Psi_{03}(\gamma,\bm r_1,\bm r_2)$,
the triple zero at $z_1=z_2$ is split into a single zero at
$z_1=z_2$, and displaced zeroes at $z_1-z_2=\pm i\sqrt{6} z_0$.
This suggests the splitting in the pattern of zeros can be used as a tool to
study the anisotropic deformation of a quantum Hall wave function.

The other FQH state,
such as the Moore-Read state\cite{MooreRead},
the Read-Rezayi state\cite{ReadRezayi},
the Haldane-Rezayi state\cite{HaldaneRezayi}, etc., could be generalized
in this way, as long as
the corresponding wave functions could be expanded in the non-interacting LLL
basis.
In addition, we also conclude that the anisotropic ${\rm Z}_k$
parafermion states \cite{ReadRezayi}
are also the (minimum total generalized angular momentum) zero-energy ground state of
some special $(k+1)$-body interaction. This kind of interaction could be expressed
in terms of the generalized $(k+1)$-body pseudopotentials \cite{Simon}
since the ${\rm Z}_k$ parafermion states are
the (minimum total angular momentum)
zero-energy ground state of $(k+1)$-body short-range interaction
\cite{Greiter,ReadRezayi}.

For example, the $\nu=1/q$ anisotropic Moore-Read states are given by
\begin{eqnarray}
\Psi_{MR}^q(\gamma)
&=&\prod_{i}\psi_{00}(\gamma,{\bm r}_i)
{\rm Pf}\left(\frac{1}{z_i-z_j+z_0^2(\partial_{z_i}-\partial_{z_j})}
\right)\nonumber\\
&&\prod_{i<j}\left[
z_i-z_j+z_0^2(\partial_{z_i}-\partial_{z_j})\right]^q 1,
\end{eqnarray}
where $\rm Pf$ means the Pfaffian, the square root of the determinant.
Among these states, the anisotropic
$\nu=1$ bosonic Moore-Read states $\Psi_{MR}^{q=1}(\gamma)$ is the
(minimum total $\bar L(\gamma)$) zero-energy ground state of
the anisotropic $3$-body pseudopotential Hamiltonian,
\begin{eqnarray}
H_{\rm int}(\gamma)=V_0h_0(\gamma),\quad V_0>0,
\end{eqnarray}
where the projection operator $h_{0}(\gamma)$ is given by
\begin{eqnarray}
&&\sum_{M=0}^\infty\sum_{m_1m_2m_3m_4m_5m_6}
{}_\gamma\langle m_1,m_2,m_3|M\rangle_\gamma\nonumber\\&&
{}_\gamma\langle M|m_4,m_5,m_6\rangle_\gamma
g_{m_1}^\dag g_{m_2}^\dag g_{m_3}^\dag
g_{m_6}g_{m_5}g_{m_4}\nonumber
\end{eqnarray}
with the $3$-body basis state $|m_1,m_2,m_3\rangle_\gamma=
|\psi_{0,m_1}(\gamma)\rangle\otimes|\psi_{0,m_2}(\gamma)\rangle
\otimes|\psi_{0,m_3}(\gamma)\rangle$
and
\begin{eqnarray}
|M\rangle_\gamma=\frac{\left({a_\gamma^\dag}_{1}+{a_\gamma^\dag}_{2}
+{a_\gamma^\dag}_{3}\right)^{M}}{\sqrt{3^M M!}}|0,0,0\rangle_\gamma
\nonumber
\end{eqnarray}
And the anisotropic $\nu=1/2$ fermionic Moore-Read states $\Psi_{MR}^{q=2}(\gamma)$ is the
(minimum $\bar L(\gamma)$) zero-energy ground state of
the following anisotropic $3$-body pseudopotential Hamiltonian,
\begin{eqnarray}
H_{\rm int}(\gamma)=V_1h_1(\gamma),\quad V_1>0,
\end{eqnarray}
where the projection operator $h_{1}(\gamma)$ is given by
\begin{eqnarray}
&&\sum_{M=0}^\infty\sum_{m_1m_2m_3m_4m_5m_6}
{}_\gamma\langle m_1,m_2,m_3|3,M-3\rangle_\gamma\nonumber\\
&&
{}_\gamma\langle 3,M-3|m_4,m_5,m_6\rangle_\gamma
g_{m_1}^\dag g_{m_2}^\dag g_{m_3}^\dag
g_{m_6}g_{m_5}g_{m_4}.\nonumber
\end{eqnarray}
with the three-body state $|3,M-3\rangle_\gamma$ being
\begin{eqnarray}
|3,M-3\rangle_\gamma&=&B_M({a_\gamma^\dag}_{1}-{a_\gamma^\dag}_{2})
({a_\gamma^\dag}_{1}-{a_\gamma^\dag}_{3})
({a_\gamma^\dag}_{2}-{a_\gamma^\dag}_{3})\nonumber\\
&&\left({a_\gamma^\dag}_{1}+{a_\gamma^\dag}_{2}
+{a_\gamma^\dag}_{3}\right)^{M-3}
|0,0,0\rangle_\gamma
\nonumber
\end{eqnarray}
and $B_{M}$ being the appropriate normalization factor.

\begin{figure}[tbp]
\centering
\includegraphics[width=3.2in]{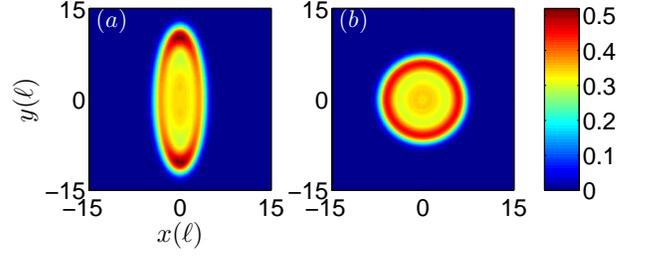}
\caption{(Color online) The density profiles
(in units of $1/(2\pi\ell^2)$) of
the $\nu=1/3$ anisotropic Laughlin state with $\gamma=1/2$
(a) and $\gamma=0$ (b) for an $N=10$ system.
Note the yellow bulk whose value approximates $1/3$.}
\label{density}
\end{figure}

\section{Numerical Results for the Anisotropic FQH states}\label{numerics}

In this section we numerically study some properties of the the
anisotropic Laughlin states $\Psi_L^q(\gamma)$ and its
applicability in a system with dipole-dipole interaction.
We first demonstrate the deformation of the FQH droplet
in density profile. The anisotropy leads also to the
deformation of the correlation hole, which can be understood by
the split of the third-order zero to three adjacent first-order
zeros. As a trivial example we show that the isotropic Laughlin
states, among the family of generalized states,
is the ground state in the variational sense for isotropic
hard-core interaction. On the other hand, the
anisotropic dipole-dipole interaction picks a variational
ground state with a finite $\gamma$ parameter as expected.

\begin{figure}[tbp]
\centering
\includegraphics[width=3.2in]{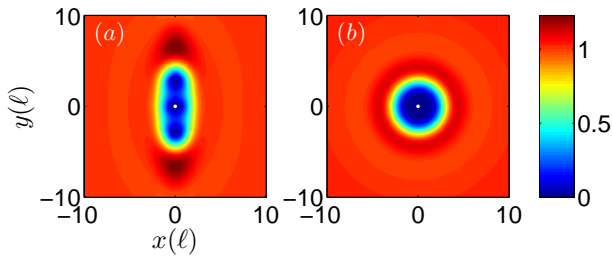}
\caption{(Color online)
The pair correlation function for $\nu=1/3$ Laughlin state
with $\gamma=1/2$
(a) and $\gamma=0$ (b) of an $N=10$ system.}
\label{pair}
\end{figure}

\subsection{Density profiles}

The density profile for the anisotropic Laughlin state is most easily
calculated using the Jack polynomials\cite{JacksFQHS,JacksF}. Explicitly,
we can write
\begin{eqnarray}
{\varrho}(\gamma,{\bm r})=\langle \Psi^q_{L}(\gamma)|
{\hat \Psi}^\dag({\bm r}){\hat \Psi}({\bm r})
|\Psi^q_{L}(\gamma)\rangle
\end{eqnarray}
for a finite number of particles.
In Fig.~\ref{density}, we plot the density profile
${\varrho}(\gamma,z)$ with $\gamma=1/2$ for an $N=10$
fermionic system at $\nu = 1/3$
and compare with that of the isotropic Laughlin wave function.
Roughly speaking, in the anisotropic case the FQH droplet
is stretched along the $y$-direction, while it maintains
its value around $1/3$, in units of $1/ (2\pi\ell^2)$ in the
bulk (indicated by the yellow color in the color plot).

\subsection{Pair correlation function}
The density-density correlation function represents the conditional
probability of find one particle at ${\bm r}$ when another is simultaneously at ${\bm r}'$.
For the anisotropic Laughlin state $\Psi_{L}^{q}(\gamma)$,
the density-density correlation function is defined by
\begin{eqnarray}
G^{(2)}(\gamma,{\bm r},{\bm r}')=\langle \Psi_{L}^{q}(\gamma)|{\hat \Psi}^\dag({\bm r})
{\hat \Psi}^\dag({\bm r}'){\hat \Psi}({\bm r}'){\hat \Psi}({\bm r})
|\Psi_{L}^{q}(\gamma).\nonumber
\end{eqnarray}
Without loss of generality, we consider the pair correlation function
with ${\bm r}'$ fixed at the origin,
\begin{eqnarray}
g(\gamma,{\bm r})=
\frac{G^{(2)}(\gamma,{\bm r},0)}{{\varrho}(\gamma,{\bm r}){\varrho}(\gamma,0)}.
\end{eqnarray}

\begin{figure}[tbp]
\centering
\includegraphics[width=3.2in]{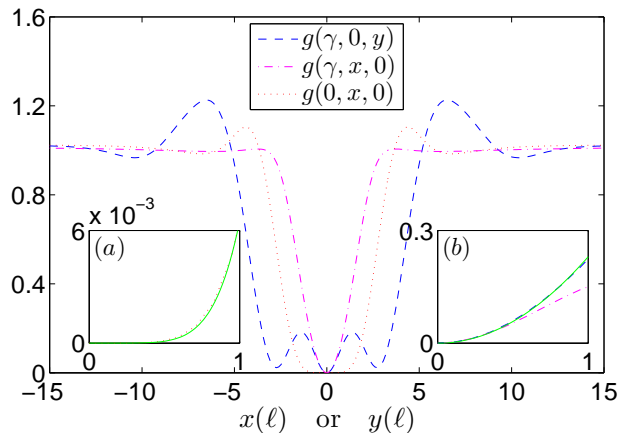}
\caption{(Color online)
The pair correlation function $g(\gamma,0,y)$, $g(\gamma,x,0)$
with $\gamma=1/2$ and $g(0,x,0)$ of an $N=10$ anisotropic Laughlin state.
The inset (a) shows the asymptotic (green solid) line
of $g(0,x,0)$, which is proportional to $x^6$.
And the inset (b) shows the asymptotic (green solid) line of
$g(\gamma,x,0)$ ($g(\gamma,0,y)$),
which is proportional to $x^2$ ($y^2$).}
\label{linepair}
\end{figure}

In Fig.~\ref{pair}, we compare $g(\gamma,{\bm r})$ of an $N=10$
Laughlin state at $\nu = 1/3$ with $\gamma = 1/2$ and 0.
The two cases are clearly distinguishable,
as the correlation hole for $\gamma = 1/2$ is stretched
{\em non-monotonically} along the $y$-direction, along which
the density profile is stretched.
The center of the hole is strictly zero but two other minima
developed in the $y$-direction are not.
To explore more details we analyze $g(\gamma,x,y)$
as a function of $x$ (or $y$) along the $y$-axis ($x$-axis)
in Fig.~\ref{linepair}.
The comparison shows that the asymptotic behavior of
$g(\gamma\neq0,{\bm r}\rightarrow0)$
is very different from that of $g(\gamma=0,{\bm r}\rightarrow0)$.
The difference roots in the toy model for two particles we
studied in Subsec.~\ref{AFQH}, from which we expect that
$g({\gamma\neq0},{\bm r}\rightarrow0)$ vanishes as $|{\bm r}|^2$,
while $g({\gamma=0},{\bm r}\rightarrow0)$ as $|{\bm r}|^6$, as indicated by
Eqs.~(\ref{2Ls}) and (\ref{a2Ls}).
The insets (a) and (b) in Fig.~\ref{linepair} confirm
the asymptotic behavior.

We emphasize that in the anisotropic case the pair correlation
function is isotropic for small enough $|{\bm r}|$, reflecting the first-order zero
in the fermionic wave function enforced by the Pauli exclusion
principle. The two-particle wave function can be regarded as the
asymptotic wave function when two particles are significantly
closer than their distance to any other particles.
For $|{\bm r}|$ comparable to $|\gamma|$, the pair correlation function
becomes anisotropic. The observed anisotropy at short distances encodes the
geometrical deformation of the Laughlin state.
When two particles are close to each other, each particle sees
$q$ zeros (for the Laughlin state at $\nu = 1/q$); the spatial spread
of them reflects the extent of the deformation. Topological properties,
manifested in the isotropic case by the $q$th-order zero, can
be identified when one look not too close ($|{\bm r}| \gg |\gamma|$).

\subsection{Variational energy}

Given the family of Laughlin states $\Psi_L^q(\gamma)$, we test
the variational principle on a trivial Hamiltonian with the
isotropic hard-core interaction, which renders the isotropic
Laughlin state as its exact zero-energy ground state, and a
Hamiltonian with anisotropic dipolar-dipolar interaction.
In both cases the expectation value of the Hamiltonian develops
a minimum, which may be identified as the variational ground
state energy. The minimum occurs at $\gamma = 0$ for the
isotropic interaction and a nonzero $\gamma$ for the
anisotropic interaction.

\begin{figure}[tbp]
\centering
\includegraphics[width=3.2in]{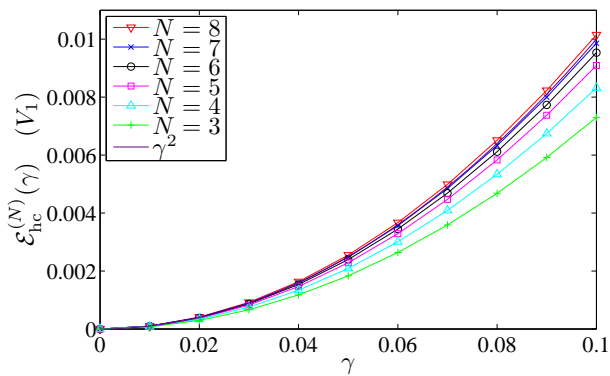}
\caption{(Color online) The hard-core interaction energy per particle
${\cal E}_{\rm hc}^{(N)}(\gamma)$ with respect to the distort parameter
$\gamma$ for $N=3-8$ system. Here we also plot $\gamma^2$
(solid line) for comparison.}
\label{eint_fvhc}
\end{figure}

Because the Laughlin wave function is the zero-energy ground state
of the (isotropic) hard-core interaction, $\gamma=0$ is naturally
the minimum for the energy expectation value. The variational energy
per particle is expected to increase as $\beta |\gamma|^2$ for
small $|\gamma|$, as can be quickly understood, e.g., from the
expansion of $\Psi^q_L(\gamma)$ with respect to small $|\gamma|$,
\begin{eqnarray}
\Psi^q_L(\gamma)=\left[1-\textstyle{\frac{1}{2}}
\gamma\sum_iz_i^2+\textstyle{\frac{1}{2}}\gamma^*\sum_i\partial_{z_i}^2+\ldots
\right]\Psi^q_L(0).
\end{eqnarray}
Here $\beta$, like stiffness, quantifies the energy cost of the
fluctuation of metric. In Fig.~\ref{eint_fvhc} we plot
the hard-core interaction energy per particle for the anisotropic fermionic
Laughlin state $\Psi^{q=3}_{L}(\gamma)$
\begin{eqnarray}
{\cal E}_{\rm hc}^{(N)}(\gamma)
&=&\frac{\langle\Psi^q_{L}(\gamma)|H_{\rm int}(\{V_m\},\gamma=0)
|\Psi^q_{L}(\gamma)\rangle}{N},
\end{eqnarray}
with $V_m=V_1\delta_{m,1}$.
For the set of finite-size systems with $N=3$-$8$,
we confirm that the minimum of the hard-core interaction energy
occurs identically at $\gamma=0$. Therefore,
the (isotropic) Laughlin wave function is indeed the optimal state for
the hard-core interaction.

\begin{figure}[tbp]
\centering
\includegraphics[width=3.2in]{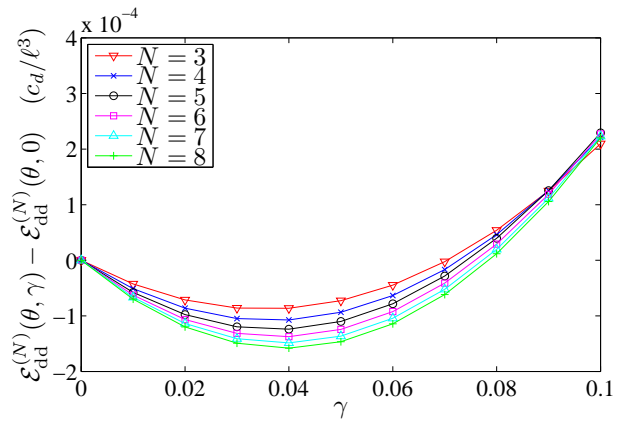}
\caption{(Color online) The dipole-dipole interaction energy per particle
${\cal E}_{\rm dd}^{(N)}(\theta=30^\circ,\gamma)$
with respect to the distort parameter
$\gamma$ for $N=3-8$ system. Here
axial oscillator length $d$ is set as $0.01\ell$.}
\label{eint_fvdd}
\end{figure}

We now turn to the dipole-dipole interaction, which some of us studied
in Ref.~\onlinecite{QHEanis} by exact diagonalization.
For the dipole-dipole interaction with
the dipole moments being polarized in the $x$-$z$ plane,
the potential in the $x$-$y$ plane is given by
\begin{eqnarray}
V(\theta,x,y)&=&\int d\xi\frac{e^{-\xi^2/2d^2}}{\sqrt{2\pi d^2}}
\frac{c_d}{(x^2+y^2+\xi^2)^{5/2}}\nonumber\\
&&\left\{x^2+y^2+\xi^2-3(\xi\cos\theta+x\sin\theta)^2\right\},\nonumber
\end{eqnarray}
where $c_d$ is the interaction strength and the motion of all the particles
along the $z$-axis is frozen to the ground state of the axial harmonic oscillator
$\pi^{-1/4}d^{-1/2}e^{-z^2/2d^2}$ with $d$ being the axial oscillator length.
Here the tilt angle $\theta$ is
introduced to tune the dipole-dipole interaction
such that $V(\theta, z)$ is isotropic (anisotropic) on $x$-$y$ plane
for $\theta=0$ ($\theta\neq0$).
The dipole-dipole interaction energy per particle in the anisotropic fermionic Laughlin state
$\Psi^{q=3}_{L}(\gamma)$ is
\begin{eqnarray}
{\cal E}_{\rm dd}^{(N)}(\theta,\gamma)
=\frac{\langle\Psi^q_{L}(\gamma)|\sum_{i<j}V(\theta,z_i-z_j)
|\Psi^q_{L}(\gamma)\rangle}{N}.
\end{eqnarray}
In Fig.~\ref{eint_fvdd}, we plot ${\cal E}_{\rm dd}^{(N)}(\theta=30^\circ,\gamma)$
with respect to $\gamma$ for finite-size system with $N=3-8$.
We found that the lowest ground state energy of the dipole-dipole interaction
occurs for the anisotropic Laughlin state with a nonzero $\gamma$.
The further quantitative comparison between the model
anisotropic quantum Hall state and the exact ground
state of the anisotropic dipole-dipole interaction will be given elsewhere.

\section{Summary}
\label{conclusion}

In this work we have explicitly constructed families of
anisotropic fractional quantum Hall states, which are exact
zero energy ground states of appropriate anisotropic short-range
two- or multi-particle interactions. These families generalize the
celebrated Laughlin, Moore-Read and Read-Rezayi states.
Each family is parameterized by a single geometric parameter
that describes the distortion of correlation hole in the
density-density correlation functions.
These states thus explicitly illustrate the existence of geometric
degree of freedom in fractional quantum Hall effect, as
recently demonstrated in Ref.~\onlinecite{Metric}.
Application of these states in studies of systems with realistic
anisotropic interactions, like dipole-dipole interaction,
will be pursued in the near future and reported elsewhere.

\section*{Acknowledgments}
This work was supported by the NSFC (Grant Nos. 11025421, 10935010 and 11174246),
the National 973 Program (Grants No. 2012CB922104 and No. 2009CB929101)
and DOE grant No. DE-SC0002140 (F.D.M.H. and K.Y.).

\end{document}